\documentclass[letter]{aa} 
\usepackage{graphicx,natbib}
\usepackage{txfonts}
\begin{document}
   \title{The high activity of 3C~454.3 in autumn 2007}

   \subtitle{Monitoring by the WEBT during the AGILE detection\thanks{The radio-to-optical 
   data presented in this paper are stored in the WEBT archive; for questions 
   regarding their availability, please contact the WEBT President Massimo 
   Villata ({\tt villata@oato.inaf.it}).}}

   \author{C.~M.~Raiteri              \inst{ 1}
   \and   M.~Villata                  \inst{ 1}
   \and   W.~P. Chen                  \inst{ 2}
   \and   W.-S. Hsiao                 \inst{ 2}
   \and   O.~M.~Kurtanidze            \inst{ 3}
   \and   K.~Nilsson                  \inst{ 4}
   \and   V.~M.~Larionov              \inst{ 5,6}
   \and   M.~A.~Gurwell               \inst{ 7}
   \and   I.~Agudo                    \inst{ 8}
   \and   H~.D.~Aller                 \inst{ 9}
   \and   M.~F.~Aller                 \inst{ 9}
   \and   E.~Angelakis                \inst{10}
   \and   A.~A.~Arkharov              \inst{ 6}
   \and   U.~Bach                     \inst{10}
   \and   M.~B\"ottcher               \inst{11}
   \and   C.~S.~Buemi                 \inst{12}
   \and   P.~Calcidese                \inst{13}
   \and   P.~Charlot                  \inst{14}
   \and   F.~D'Ammando                \inst{15}
   \and   I.~Donnarumma               \inst{15}
   \and   E.~Forn\'e                  \inst{16}
   \and   A.~Frasca                   \inst{12}
   \and   L.~Fuhrmann                 \inst{10}
   \and   J.~L.~G\'{o}mez             \inst{ 8}
   \and   V.A.~Hagen-Thorn            \inst{ 5}
   \and   S.~G.~Jorstad               \inst{17}
   \and   G.~N.~Kimeridze             \inst{ 3}
   \and   T.~P.~Krichbaum             \inst{10}
   \and   A.~L\"ahteenm\"aki          \inst{18}
   \and   L.~Lanteri                  \inst{ 1}
   \and   G.~Latev                    \inst{19}
   \and   J.-F.~Le~Campion            \inst{14}
   \and   C.-U.~Lee                   \inst{20}
   \and   P.~Leto                     \inst{21}
   \and   H.-C. Lin                   \inst{ 2}
   \and   N.~Marchili                 \inst{10}
   \and   E.~Marilli                  \inst{12}
   \and   A.~P.~Marscher              \inst{17}
   \and   R.~Nesci                    \inst{22}
   \and   E.~Nieppola                 \inst{18}
   \and   M.~G.~Nikolashvili          \inst{ 3}
   \and   J.~Ohlert                   \inst{23}
   \and   E.~Ovcharov                 \inst{19}
   \and   D.~Principe                 \inst{11}
   \and   T.~Pursimo                  \inst{24}
   \and   B.~Ragozzine                \inst{11}
   \and   A.~C.~Sadun                 \inst{25}
   \and   L.~A.~Sigua                 \inst{ 3}
   \and   R.~L.~Smart                 \inst{ 1}
   \and   A.~Strigachev               \inst{26}
   \and   L.~O.~Takalo                \inst{ 4}
   \and   M.~Tavani                   \inst{15}
   \and   C.~Thum                     \inst{27}
   \and   M.~Tornikoski               \inst{18}
   \and   C.~Trigilio                 \inst{12}
   \and   K.~Uckert                   \inst{11}
   \and   G.~Umana                    \inst{12}
   \and   A.~Valcheva                 \inst{26}
   \and   S.~Vercellone               \inst{28}
   \and   A.~Volvach                  \inst{29}
   \and   H.~Wiesemeyer               \inst{30}
 }

   \offprints{C.~M.~Raiteri}

   \institute{
          INAF, Osservatorio Astronomico di Torino, Italy                                                     
   \and   Institute of Astronomy, National Central University, Taiwan                                         
   \and   Abastumani Astrophysical Observatory, Georgia                                                       
   \and   Tuorla Observatory, Univ.\ of Turku, Piikki\"{o}, Finland                                           
   \and   Astron.\ Inst., St.-Petersburg State Univ., Russia                                                  
   \and   Pulkovo Observatory, St.\ Petersburg, Russia                                                        
   \and   Harvard-Smithsonian Center for Astroph., Cambridge, MA, USA                                         
   \and   Instituto de Astrof\'{i}sica de Andaluc\'{i}a (CSIC), Granada, Spain                                
   \and   Department of Astronomy, University of Michigan, MI, USA                                            
   \and   Max-Planck-Institut f\"ur Radioastronomie, Bonn, Germany                                            
   \and   Department of Physics and Astronomy, Ohio Univ., OH, USA                                            
   \and   INAF, Osservatorio Astrofisico di Catania, Italy                                                    
   \and   Osservatorio Astronomico della Regione Autonoma Valle d'Aosta, Italy                                
   \and   Lab.\ d'Astrophysique, Universit\'e Bordeaux 1, CNRS, Floirac, France                               
   \and   INAF, IASF-Roma, Italy                                                                              
   \and   Agrupaci\'o Astron\`omica de Sabadell, Spain                                                        
   \and   Institute for Astrophysical Research, Boston University, MA, USA                                    
   \and   Mets\"ahovi Radio Obs., Helsinki Univ.\ of Technology TKK, Finland                                  
   \and   Sofia University, Bulgaria                                                                          
   \and   Korea Astronomy and Space Science Institute, South Korea                                            
   \and   INAF, Istituto di Radioastronomia, Sezione di Noto, Italy                                           
   \and   Dept.\ of Phys.\ ``La Sapienza" Univ, Roma, Italy                                                   
   \and   Michael Adrian Observatory, Trebur, Germany                                                         
   \and   Nordic Optical Telescope, Santa Cruz de La Palma, Spain                                             
   \and   Dept.\ of Phys., Univ.\ of Colorado, Denver, USA                                                    
   \and   Inst.\ of Astronomy, Bulgarian Academy of Sciences, Sofia, Bulgaria                                 
   \and   Institut de Radio Astronomie Millim\'etrique, Grenoble, France                                      
   \and   INAF, IASF-Milano, Italy                                                                            
   \and   Radio Astronomy Lab.\ of Crimean Astrophysical Observatory, Ukraine                                 
   \and   Instituto de Radioastronom\'{i}a Millim\'{e}trica, Granada, Spain                                   
 }

   \date{}

 
  \abstract
   {The quasar-type blazar \object{3C 454.3} underwent a phase of high activity in summer and autumn 2007,
which was intensively monitored in the radio-to-optical bands by the Whole Earth Blazar Telescope 
(WEBT). The $\gamma$-ray satellite Astro-rivelatore Gamma a Immagini LEggero (AGILE) detected 
this source first in late July, and then in November--December 2007.} 
   {In this letter we present the multifrequency data collected by the WEBT and collaborators
during the second AGILE observing period, complemented by a few contemporaneous data from
the UltraViolet and Optical Telescope (UVOT) onboard the Swift satellite.
The aim is to trace in detail the behaviour of the synchrotron emission from the blazar jet, 
and to investigate the contribution from the thermal emission component.}
   {Optical data from about twenty telescopes have been homogeneously calibrated and carefully assembled 
to construct an $R$-band light curve containing about 1340 data points in 42 days. 
This extremely well-sampled optical light curve allows us to follow the dramatic 
flux variability of the source in detail. In addition, we show radio-to-UV spectral energy distributions (SEDs) at different epochs, which represent different brightness levels.}
   {In the considered period, the source varied by 2.6 mag in a couple of weeks in the $R$ band. 
Many episodes of fast (i.e.\ intranight) variability were observed, most notably on December 12,
when a flux increase of about 1.1 mag in 1.5 hours was detected, 
followed by a steep decrease of about 1.2 mag in 1 hour.
The contribution by the thermal component is difficult to assess, due to the uncertainties in the Galactic,
and possibly also intrinsic, extinction in the UV band. However, polynomial fitting of radio-to-UV SEDs reveals an increasing spectral bending going towards fainter states, suggesting a UV excess likely due to the thermal emission from the accretion disc.}
  {Once the AGILE data are completely analysed, the low-frequency observations presented in this letter 
will offer a formidable tool to investigate the optical-$\gamma$ flux correlations, i.e.\ the relationship between the synchrotron and inverse-Compton emission components.}



   \maketitle
%

\section{Introduction}

The flat-spectrum radio quasar \object{3C 454.3} focused the attention of astronomers in spring 2005,
when it was observed in an exceptional brightness state from the near-IR to the X-ray frequency range.
The Whole Earth Blazar Telescope (WEBT)\footnote{{\tt http://www.oato.inaf.it/blazars/webt/} \\ see e.g.\ \citet{vil04b,boe07,rai08}.} organised a multiwavelength campaign to study the event, whose results were published by \cite{vil06}.
The WEBT monitoring effort continued after the optical outburst, and followed the subsequent radio activity \citep{vil07}, and then the faint state in the 2006--2007 observing season \citep{rai07b}.
In this last period, the low contribution by the synchrotron emission from the jet allowed the authors to recognise both the little blue bump, due to line emission from the broad line region, and the big blue bump, due to thermal emission from the accretion disc.

A renewed optical activity was observed at the beginning of the next observing season, in May 2007, which prompted the WEBT to go on with the monitoring. 
Indeed, a big optical outburst was observed in July--August.
This triggered observations by the Astro-rivelatore Gamma a Immagini LEggero (AGILE) satellite, which detected the source in its brightest $\gamma$-ray state ever observed \citep{ver08}. A further significant detection by AGILE was announced in November \citep{che07,puc07}, which in turn led to intensified monitoring by the WEBT.

The importance of collecting data at both low and high energies simultaneously comes from the fact that the  X-to-$\gamma$-ray radiation from blazars is commonly thought to be produced in the plasma jet through an inverse-Compton process off the same relativistic electrons that generate the radio-to-optical synchrotron emission. The seed photons for the inverse-Compton scattering may be either the same synchrotron photons (synchrotron-self-Compton or SSC models) or photons coming from outside the jet (external-Compton or EC models), or a combination of both. In particular, the $\gamma$-ray radiation in the GeV regime should be directly connected to the optical one. Hence, the comparison between the $\gamma$-ray AGILE data and the optical data, especially during periods of high variability, can give an important contribution to our knowledge of blazar emission mechanisms.
An example of synergy between AGILE and WEBT has recently been put into practice on the blazar S5 0716+71 \citep{vil08}.

In this letter we deal with the multiwavelength data taken in the core period (November 4--December 16, 2007)
of the last WEBT campaign on 3C~454.3. This core period corresponds to the AGILE second detection, and includes a few observations by the Swift satellite.
We will present a detailed analysis of all multiwavelength data acquired during 
the whole last WEBT campaign (May 2007--February 2008) in a forthcoming paper (Raiteri et al., in prep.).

\section{Optical-to-radio observations by the WEBT}

Optical and near-IR data were acquired in Johnson-Cousins $UBVRI$ and $JHK$ bands by about twenty telescopes all around the world.
Magnitudes were calibrated following \cite{ang71} for the $U$ filter, \cite{rai98} for the $BVR$ filters, and \cite{gon01} for the $I$ filter. The $JHK$ data were calibrated according to a photometric sequence by Larionov (private commun.).
The light curves construction was performed by assembling all datasets for each band and then simultaneously inspecting all the resulting optical--NIR light curves day by day. When an intranight dataset contained noisy data, they were binned (typically on 5 minutes); redundant data points with too large errors ($\ga 0.1$ mag) and unreliable outliers were discarded. 

\begin{figure}
   \centering
   \resizebox{\hsize}{!}{\includegraphics{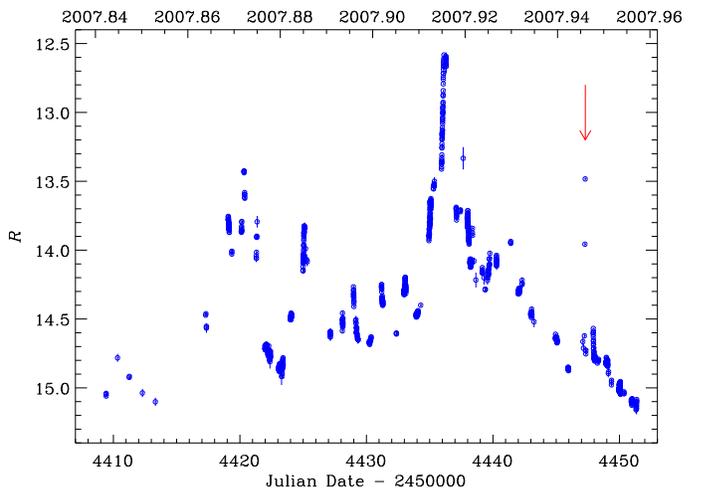}}
      \caption{$R$-band light curve of 3C~454.3 in November 4--December 16, 2007. 
The episode of extremely fast variability indicated by the red arrow is shown in more detail in Fig.\ \ref{dec12}.} 
         \label{rband}
   \end{figure}

The $R$-band light curve obtained during the core period of the WEBT campaign is shown in Fig.\ \ref{rband}. It was constructed with the data provided by the following observatories (ordered by decreasing contribution): 
Lulin, 
Abastumani, 
Roque de los Muchachos (KVA and NOT),
St.\ Petersburg, 
Mt.\ Lemmon,
Valle d'Aosta,
Torino,
Kitt Peak (MDM),
L'Ampolla,
Vallinfreda,
Michael Adrian,
Rozhen (50/70 cm and 200 cm telescopes),
New Mexico Skies, and
Calar Alto\footnote{Calar Alto data were taken as part of the MAPCAT (Monitoring AGN with Polarimetry at the Calar Alto Telescopes) project.}.
The light curve contains 1337 data points, which show a total variation of 2.58 mag, occurred in the last 15.18 days.
The Bordeaux and Catania observatories also contributed to the optical light curves, but not in the $R$ band.

\begin{figure}
   \centering
   \resizebox{\hsize}{!}{\includegraphics{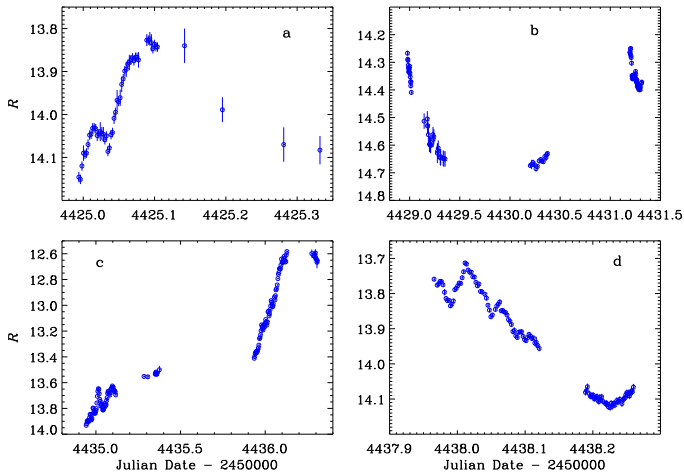}}
      \caption{Examples of very fast variability of 3C~454.3 extracted from the light curve in Fig.\ \ref{rband}.} 
         \label{idv}
   \end{figure}

The source was very active in this period, and strong variations were observed also on intranight 
time scales. Noticeable episodes of fast variability are displayed in Fig.\ \ref{idv}.
Figure 2a shows a well-sampled brightening of 0.33 mag in 2.3 hours, which was followed by an 
apparently slower flux decrease. 
The first dimming phase in Fig.\ 2b implies a variation of 0.38 mag in 9.0 hours.
A very spectacular variability episode is shown in Fig.\ 2c, where the total flux increase is 1.35 mag in 28.5 hours, including an almost monotonic brightening of 0.83 mag in 4.6 hours\footnote{With very steep slopes of about 0.006 mag per minute, i.e.\ 3 times steeper than the fastest variations observed in
the strong IDV blazar 0716+714 \citep[see][and references therein]{vil08}.},
from JD = 2454435.9357 to 2454436.1284, corresponding to the maximum peak of the light curve in Fig.\ \ref{rband}. 
Finally, the flux variations displayed in Fig.\ 2d involve a 0.41 mag dimming in 5.1 hours.

\begin{figure}
   \centering
   \resizebox{\hsize}{!}{\includegraphics{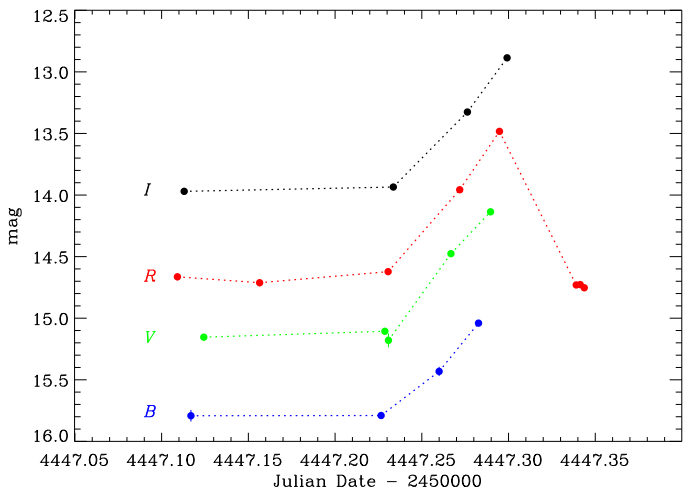}}
      \caption{The extraordinary episode of fast variability observed on December 12, 2007.} 
         \label{dec12}
   \end{figure}

A special mention is due to the episode of extraordinarily fast variability detected on December 12, 2007, which is shown in Fig.\ \ref{dec12}. Observations started in St.\ Petersburg ($BVRI$) and then continued in Valle d'Aosta ($BVRI$) and Bordeaux ($V$). 
For all these datasets the source was relatively faint. But soon after,
eight $BVRI$ frames taken at the Torino Observatory revealed a sudden brightening, which was already over when the Roque (KVA) $R$-band data were acquired. This rapid flare implies at least a 1.1 mag brightening
in 1.5 hours, and a brightness decrease of about 1.2 mag in 1 hour in the $R$ band\footnote{More precisely,
1.25 mag in 64 minutes, i.e.\ a mean slope of 0.020 mag per minute (cf.\ Footnote 3).}, thus representing one of the most dramatic flux changes ever detected in blazars.
For this reason the Torino frames were carefully analysed to check for any possible instrumental problem and artifact, finding none. Hence, we believe that the observed fast variation represents a real phenomenon. 

At the beginning of the period considered in this letter, a few near-IR data were taken at Campo Imperatore and with the Nordic Optical Telescope (NOT).

Radio data at both mm and cm wavelengths were collected as already calibrated flux densities.
In the mm regime, data were taken at 850 $\mu \rm m$ and 1 mm with the 
SubMillimiter Array (SMA) on Mauna Kea\footnote{These data were obtained as part of the normal monitoring program initiated by the SMA \citep{gur07}.}, and at 1 mm, 142 GHz, and 86 GHz with the IRAM 30 m telescope on Pico Veleta\footnote{The IRAM data are part of a standard AGN monitoring program of total flux and polarization, a complementary program to monitor additional sources selected for observations with GLAST (Fuhrmann et al., in prep.), and a dedicated polarimetry program to monitor the mm flare of 3C~454.3 in 2007 (Agudo et al., in prep.).}.
At cm wavelengths, the radio flux densities were measured at the 
Mets\"ahovi (37 GHz), 
UMRAO (14.5, 8.0, and 4.8 GHz),
Effelsberg (42, 23, 15, 8, 5, 3, and 1 GHz), 
Crimean (37 GHz),
Medicina (22, 8, and 5 GHz), and
Noto (43 GHz) observatories.
We also collected data from the VLA/VLBA Polarization Calibration Database\footnote{\tt http://www.vla.nrao.edu/astro/calib/polar/}.

\section{UV data from Swift and spectral energy distributions}

In the period considered in this letter, observations by the UVOT instrument \citep{rom05} onboard Swift 
were performed on November 15, 20, and 22, and on December 13 and 15.
However, only in the first and in the last two epochs all the optical ($UBV$) and UV ($W2, M2, W1$) filters were used, while in the other epochs data in the UV bands only were acquired.
The UVOT data were processed with the {\tt uvotmaghist} task of the HEASOFT 6.3 package.
Following the recommendations contained in the release notes, the source counts were extracted from a circular region with a 5 arcsec radius, in agreement with the standard photometric aperture defined in the calibration files (CALDB updated as July 2007). 
To avoid contamination by Star 1 \citep[see][]{rai98}, the background was extracted from source-free circular 
regions in the source surroundings.
We attributed a 0.1 mag uncertainty to all UVOT data, to take account of both systematic and statistical errors.

\begin{figure}
   \centering
   \resizebox{\hsize}{!}{\includegraphics{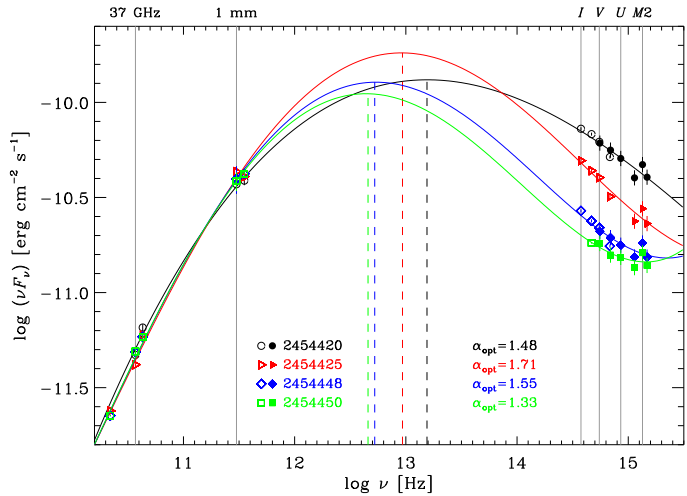}}
      \caption{Spectral energy distributions (SEDs) of 3C~454.3 at different brightness levels in November--December 2007. Contemporaneous data are plotted with the same symbol and colour; filled symbols refer to Swift-UVOT data,
while empty symbols represent ground-based observations. The curves show polynomial fits to the radio-to-UV data at the various epochs; dashed vertical lines correspond to the fit maxima. The values of the optical spectral indices are also indicated.} 
         \label{sed}
   \end{figure}

We constructed broad-band SEDs with contemporaneous data from the radio bands to the UV frequencies.
We selected epochs in which data at various frequencies were available, to better trace the synchrotron component. In Fig.\ \ref{sed} we show the most significant SEDs at different brightness levels.
Before being converted into flux densities, NIR-to-UV magnitudes were de-reddened according to the \cite{car89} laws,
assuming an extinction in the $B$ band of 0.462 mag, and $R_V=A(V)/E(B-V)=3.1$, as in the diffuse interstellar medium. One delicate point is the amount of extinction in the UV bands.
The UVOT effective wavelengths for the UV$W1$, UV$M2$, and UV$W2$ filters are 2634, 2231, and 2030 \AA, respectively \citep{poo08}. We derive for these bands an extinction of 0.73, 1.07, and 1.02 mag, respectively, and the result is the wavy shape of the SEDs in Fig.\ \ref{sed}.
We notice that the amount of Galactic extinction in the UV is strongly affected
by the 2175 \AA\ bump due to dust absorption. Even slight variations in the depth and/or shape of this feature would have noticeable consequences on the SEDs in the UV, particularly on the position of the UV$M2$ points, since the effective wavelength of this filter is close to 2175 \AA.
Moreover, we remind that fitting the 3C~454.3 X-ray spectra requires extra absorption \citep[see e.g.][and references therein]{rai07b}. Hence, there may be intrinsic absorption also in the optical--UV frequency range \citep[see e.g.][]{gas04}, even if it may have different properties with respect to the Galactic extinction \citep{mai01}.

Consequently, the shape of the spectrum in the UV band has to be taken with caution.
However, if we perform a polynomial fit of the SEDs from the radio (starting from 5 GHz, outside the plot in Fig.\ \ref{sed}) to the UV data, we can see that the optical--UV SED becomes more and more concave as the flux decreases, suggesting the presence of a rather stable UV excess.
Indeed, a UV excess was found by \citet{rai07b} when analysing XMM-Newton data taken in July and December 2006, which was interpreted as the contribution from the accretion disc. Our SEDs would confirm the picture according to which the thermal contribution from the disc is overwhelmed by the beamed synchrotron emission when the source is bright, and it can be recognised only when the non-thermal flux is low.

We finally notice the remarkable change in the optical spectral index $\alpha_{\rm opt}$ ($F_\nu \propto \nu^{- \alpha}$), from 1.33 (JD = 2454450) to 1.71 (JD = 2454425), and that the polynomial fits suggest that the peak of the SED may shift towards higher frequencies with increasing optical flux (see Fig.\ \ref{sed}).

\section{Discussion and conclusions}

In this letter we have presented the radio-to-optical data on 3C~454.3 acquired by the WEBT and collaborators,
complemented by data taken with the UVOT instrument onboard Swift, during the core period of the last WEBT campaign, i.e.\ from November 4 to December 16, 2007.
The source showed strong activity in the optical band, with a maximum variation of about 2.6 mag in a couple of weeks. 
Many episodes of fast variability were observed, involving flux changes of several tenths of mag in 
a few hours. The most striking event was detected on December 12, when a flux increase of about 1.1 mag in 1.5 hours was followed by a fall of about 1.2 mag in 1 hour. This is one of the fastest variations ever observed in blazars, even if 
the inferred brightness temperature is still well below the limit of $\sim 10^{12} \rm \, K$ for inverse-Compton catastrophe. 
Indeed, by assuming $H_0=71 \rm \, km \, s^{-1} \, Mpc^{-1}$, 
the 1.2 mag dimming in 1 hour in the $R$ band yields  $T_{\rm b} \sim 5 \times 10^{9} \rm \, K$ 
\citep[e.g.][]{wag96}\footnote{ For comparison, the 0.83 mag brightening in 4.6 hours observed on December 1 
(JD $\sim$ 2454436.0, see Fig.\ \ref{idv}) implies $T_{\rm b} \sim 4 \times 10^{8} \rm \, K$.}.
We think that this very fast flare is most likely a real variation in the jet emission.
Alternatively, we may think of a dramatic event in the host galaxy, like a $\gamma$-ray burst (GRB). However, most GRBs are associated with the death of young massive stars, which are more common in star-forming galaxies \citep{sav08}, while blazars are usually hosted in ellipticals. 
A microlensing effect, e.g.\ by a MACHO in our Galaxy, may be possible, but it seems unlikely; indeed, these events are observed with much longer time scales \citep[e.g.][]{woo05}.
Unfortunately, the poor sampling prevents us from any consideration on the shape and chromatism of the event.

The construction of broad-band SEDs with contemporaneous data reveals noticeable spectral changes in the optical--UV frequency range. Although the details of the spectral shape must be regarded with caution because of the uncertainties in the modelling of absorption (Galactic, but possibly also intrinsic), especially in the UV band, we notice that the 
optical--UV SED gets more and more concave as the flux decreases.
This is consistent with the finding of a UV excess in faint states \citep{rai07b}, likely due to thermal radiation from the accretion disc. Indeed, this contribution to the blazar emission becomes more visible when the beamed synchrotron radiation from the jet is fainter, which stresses the importance of studying blazars in low-activity states as well as during outbursts.

The time interval considered in this letter corresponds to the second period of detection in the $\gamma$-rays by the AGILE satellite, which follows the observation of an unprecedented bright $\gamma$-ray state in July 2007 \citep{ver08}.
The AGILE data in November--December 2007 show high variability on time scales from 
days to weeks, with recurrent flaring activity (Vercellone et al., in prep.; Donnarumma 
et al., in prep.).
The comparison between the low-energy data presented in this letter and the high-energy data by AGILE will contribute to shed light on the relationship between the synchrotron and inverse-Compton emission components from the blazar jet.

\begin{acknowledgements}
We acknowledge the use of public data from the Swift data archive.
This research has made use of data obtained through the High Energy Astrophysics Science Archive Research Center Online Service, provided by the NASA/Goddard Space Flight Center.
This work is partly based on observations made with the Nordic Optical Telescope, operated
on the island of La Palma jointly by Denmark, Finland, Iceland,
Norway, and Sweden, in the Spanish Observatorio del Roque de los
Muchachos of the Instituto de Astrofisica de Canarias, and on observations collected at the German-Spanish Calar Alto Observatory, jointly operated by the MPIA and the IAA-CSIC.
AZT-24 observations are made within an agreement between  Pulkovo, Rome and Teramo observatories.
We thank the staff at the IRAM 30-m telescope for their help with the observations, and in particular H. Ungerechts for allowing us to include data from the general IRAM 30-meter AGN monitoring program. We are also grateful to the IRAM Director for providing discretionary observing time at the 30-m telescope. This paper is based on observations carried out at the IRAM 30-m telescope. IRAM is supported by INSU/CNRS (France), MPG (Germany) and IGN (Spain).
This research has made use of data from the University of Michigan Radio Astronomy Observatory,
which is supported by the National Science Foundation and NASA and by funds from the University of Michigan.
The Mets\"ahovi team acknowledges the support from the Academy of Finland.
The Torino team acknowledges financial support by the Italian Space Agency through contract 
ASI-INAF I/088/06/0 for the Study of High-Energy Astrophysics. 
Acquisition of the MAPCAT data is supported in part by the Spanish Ministerio de Educacion y Ciencia through grant AYA2007-67626-C03-03
\end{acknowledgements}

\end{document}